# Valley-spin blockade and spin resonance in carbon nanotubes

Fei Pei, Edward A. Laird, Gary A. Steele & Leo P. Kouwenhoven*

*To whom correspondence should be addressed. E-mail: l.p.kouwenhoven@tudelft.nl

Kavli Institute of NanoScience, Delft University of Technology,
PO Box 5046, 2600 GA, Delft, The Netherlands

**Manipulation and readout of spin qubits in quantum dots made in III-V materials successfully rely on Pauli blockade that forbids transitions between spin-triplet and spin-singlet states[1-5]. Quantum dots in group IV materials have the advantage of avoiding decoherence from the hyperfine interaction by purifying them with only zero-spin nuclei[6]. Complications of group IV materials arise from the valley degeneracies in the electronic bandstructure. These lead to complicated multiplet states even for two-electron quantum dots thereby significantly weakening the selection rules for Pauli blockade. Only recently have spin qubits been realized in silicon devices where the valley degeneracy is lifted by strain and spatial confinement[7]. In carbon nanotubes Pauli blockade can be observed by lifting valley degeneracy through disorder[8-11]. In clean nanotubes, quantum dots have to be made ultra-small to obtain a large energy difference between the relevant multiplet states[12,13]. Here we report on low-disorder nanotubes and demonstrate Pauli blockade based on both valley and spin selection rules. We exploit the bandgap of the nanotube to obtain a large level spacing and thereby a robust blockade. Single-electron spin resonance is detected using the blockade.**

Two quantum dots containing in total two electrons can be tuned to the transition between two charge states: (1,1) with one electron in each dot and (2,0) with both electrons in the first dot. The transition involves the tunnelling of the electron from the second to the first dot. Even when this transition is allowed energetically, it can be blocked by selection rules[14]. In III-V quantum dots (e.g. GaAs or InAs) a blockade can be set up between a (1,1)-triplet state and a (2,0)-singlet state. Important for a robust blockade is the condition that the (2,0)-triplet state is high up in energy, since this excited state is not blocked by selection rules. The crucial energy difference, $E_{ST}$, between the (2,0)-triplet and (2,0)-singlet states, is several meV's in the III-V materials[14]. In carbon nanotubes the two-electron states are grouped into singlet-like and triplet-like states. The energy difference, $E'_{ST}$, between the two classes can be one or two orders of magnitude smaller for two main reasons: additional levels from valley degeneracy and stronger Coulomb interactions[12,13,15]. These complications have prevented a consistent observation of Pauli blockade and as a result spin manipulation has not been realized. We avoid these complications by using the large level spacing from the bandgap of the nanotube and demonstrate a robust valley-spin blockade. First we discuss our novel fabrication method to obtain ultra-clean quantum dots controlled by a set of gate electrodes that have high-frequency bandwidth.

Two approaches are usually used to fabricate nanotube quantum dot devices: depositing contacts and gates after a nanotube is grown and located on a substrate[9,16], or growing a suspended nanotube over predefined contacts and gates[15,17,18]. The latter approach eliminates contamination of the nanotube by chemical processing, creating an ultra-clean nanotube device. However, the high growth temperature limits the choice of materials and the device design, limiting these devices to having large gate spacings and thus a quantum dot confinement insufficient to overcome Coulomb interaction effects.

To create ultra-clean double quantum dots with optimal confinement, we have developed a novel stamping technique (see Fig. 1a and Methods), following the pioneering work by Wu et al.[19]. A single nanotube is transferred from the growth chip to the device chip under ambient conditions. The thickness of the device contacts is optimized so that in most devices, the center of the nanotube touches the trench bottom, resulting in a bend (Fig. 1b), as proposed to be necessary for electrically driven spin resonance[20]. Five gate electrodes are embedded underneath the nanotube. A double dot potential is created by the combination of Schottky barriers at the contacts and voltages applied to the gates. By tuning these gate voltages, we can populate each dot with a well-defined number of electrons or holes.

The device exhibits ambipolar double quantum dot behavior as seen from the charge stability diagram of Fig. 1c, where current through the double dot is measured as a function of the outermost gate voltages. Depending on the gate voltages, we can configure the device in a p-n, n-n, n-p or p-p region and determine the exact charge occupation number of each dot. For each region, we can find Coulomb blockade features exhibiting a characteristic fourfold periodicity of addition energy in both quantum dots, indicating shell-filling of electrons and holes. The first shells of electrons and holes are separated by a 30 meV bandgap.

The key signature of Pauli blockade is a current suppression for one direction of the source-drain bias[1,14]. In our device, blockade becomes most evident with the double dot tuned into the p-n region. This is shown in Fig. 2a,b where we observe multiple triple point bias triangles with a suppressed current at the baseline depending on the bias direction. As expected, blockade is observed for transitions where in the initial configuration, both dots contain an odd number of electrons.

A robust Pauli blockade requires a large $E'_{ST}$ conventionally provided by a strong dot confinement. Here we exploit the high tunability of our double dot to obtain a particularly large $E'_{ST}$ that includes the bandgap of the nanotube (Fig. 2c). We focus on the (3h,1e) → (2h,0) transition, where both shells initially contain one electron. Taking account of both valley ($K$ or $K'$) and spin ($\uparrow$ or $\downarrow$) quantum numbers, each shell contains four states, denoted as: $K\downarrow$, $K'\uparrow$, $K\uparrow$ and $K'\downarrow$. Spin-orbit coupling splits each shell into two doublets with energy difference $\Delta_{SO}$ at zero magnetic field[18]. For negative bias (Fig. 2d,f), current flows as either of the two electrons in the left dot can tunnel to the empty shell on the right. At positive bias (Fig. 2e), assuming that valley and spin are conserved during tunnelling, current is blocked when initial and final states differ in either valley or spin quantum numbers. The blockade is lifted when the interdot energy detuning is large enough that the initial state has access to additional final states involving a higher shell. However, as the nearest higher shell in the left dot is across the bandgap, this situation does not arise with a 10 mV bias (Fig. 2g). In the case of only spin blockade, the blockade is lifted as soon as the initial state has access to an empty final state with the same spin. Spin blockade would happen if the disorder-induced valley-mixing term $\Delta_{KK'}$ is different between the left and right dot[21]. For lifting valley-spin blockade, the empty final state must have both the same valley and spin, which leads to an additional valley selection rule for interdot tunnelling. This additional selection rule leads to suppression of the current across the entire triangles in Fig. 2g. This current suppression, which in contrast to spin blockade continues even when transport occurs via excited states of the left dot, is the unambiguous signature of valley-spin blockade.

In Fig. 3, we investigate valley and spin relaxation by measuring the leakage current for different orientations of the magnetic field. We stay at the valley-spin blockaded triangles shown in Fig. 2g with detuning axis marked by the black arrow. Fig. 3a shows the leakage current as a function of detuning and magnetic field $B_z$ along the z-axis defined in Fig. 1a. Leakage current is due to valley-spin relaxation and can arise from spin-orbit interaction[6,9,20,22], intervalley scattering[23], and hyperfine interaction with the ~1% $^{13}$C lattice nuclei[9].

Three transitions, mediated by valley-spin relaxation, are identified using a two-electron double dot model (Fig. 3a,b). For simplicity, we model the charge states (3h,1e) as (1,1), (2h,0) as (2,0), and ignore the higher shells in the left dot that remain empty in our experiments. Valley (*v*) and spin (*s*) together lead to 16 two-electron states, grouped into six valley-spin antisymmetric (singlet-like (*S*)) states with both electrons in one shell, and ten symmetric (triplet-like (*T*)) states for which two shells are required [24-26]. These 16 linearly independent states are listed explicitly in Supplementary Information. We write these states with the following shortened notation $S(T)v_1 s_1 v_2 s_2 = (|v_1 s_1 v_2 s_2\rangle - (+)|v_2 s_2 v_1 s_1\rangle)n$, (*n*, normalization factor). We use the lowest energy $TK\downarrow K\downarrow(1,1)$ state as a spectroscopic probe to measure the (2,0) spectrum[27]. The measured transitions (dashed lines in Fig. 3a) are in good agreement with the calculated spectrum. The calculation incorporates two parameters: $\Delta_{SO} = 1.6$ meV measured by the difference in transition detunings at zero field (the large $\Delta_{SO}$ has been observed in multiple devices and is the subject of ongoing investigation), and orbital magnetic moment $\mu_{orb} = 0.9$ meV/T measured by the slope of the transitions with field. (Note that the slope of the transitions changes at 2.2 T, presumably because the (1,1) ground state changes at that field.) Six valley-spin relaxation transitions are possible, however only three are observed. The relaxation rates that determine which transitions are visible in the data are not fully understood. (Further discussions in Supplementary Information.)

The orbital magnetic moment pointing along the nanotube axis leads to a large *g*-factor anisotropy. When varying $B_z$ (Fig. 3a), we couple to both the orbital magnetic moment and to the Zeeman energy, and the two transitions therefore have much larger slopes compared to varying $B_y$ (Fig. 3c), in which case we only couple to the spin. Fig. 3f shows the current as a function of field angle for fixed |**B**| = 2 T. The measured transitions show excellent agreement with a model incorporating the *g*-factor anisotropy.

An interesting consequence of the level structure in nanotubes is that valley-spin blockade appears also for the initially unblockaded bias direction at finite $B_z$. When the magnetic field induces a ground state crossing from $SK{\downarrow}K'{\uparrow}(1,1)$ to $TK{\downarrow}K{\downarrow}(1,1)$ the current becomes blocked[26]. This is evident in Fig. 3d inside the region indicated by the purple circle and the corresponding levels are illustrated in Fig. 3e. A kink observed at 1T for the lowest transitions (black) arises because at this field $SK{\downarrow}K{\uparrow}$ takes over from $SK{\downarrow}K'{\uparrow}$ as the ground state in (2,0) (see Supplementary Information).

In Fig. 4, we explore the consequences for magnetotransport of the bend expected in this device. Fig. 4a shows a leakage current anisotropy when a magnetic field is rotated in the x-y plane. Current near zero detuning is more suppressed when the field is pointed along the y-axis. This can be explained by considering that at low detuning transport only proceeds via single-particle states $K{\downarrow}$ and $K'{\uparrow}$. This doublet forms an effective spin-1/2 system governed by a Zeeman field[20] $\mathbf{B}_{\text{eff}} = \mathbf{g} \cdot \mathbf{B}/2$, which due to the bend can be different in the two dots. The effective spins in the two dots precess about these different $\mathbf{B}_{\text{eff}}$ axes so that an initially parallel effective spin state acquires an antiparallel component. This causes lifting of valley-spin blockade, resulting in a higher leakage current. However, in the z-y plane the projection of the nanotube is straight. Thus when $B_y$ is applied, $\mathbf{B}_{\text{eff}}$ is the same in both dots and valley-spin blockade remains. As expected, the current is isotropic for the unblockaded bias direction (Fig. 4c).

The observation of valley-spin blockade in a bent nanotube allows detection and driving of electric dipole spin resonance (EDSR). We use a microwave-frequency signal added to $V_L$ to oscillate electrons in the double dot. Fig. 4e shows the current as a function of $B_z$ and microwave frequency at a blockaded transition in the many-electron regime of a second device. When the frequency of the microwave matches the splitting of two valley-spin states, the blockade is partly lifted. EDSR is observed as V-shaped lines with slopes yielding $g = 2$. The relatively small $g$-factor in this second device is presumably due to a large electron occupation or disorder[28]. EDSR is also detected in the first device in a different cooldown, with a rather complex spectrum shown in the Supplementary Information.

In summary, we have developed a new fabrication method to make a double quantum dot in a bent carbon nanotube using stamping technique. The devices exhibit an exceptional confinement and tunability, which enable us to observe valley-spin blockade and to demonstrate electric dipole spin resonance. Our results indicate the feasibility of valley-spin qubits in carbon nanotubes.

## Methods

Fabrication includes the carbon nanotube growth chip and the device chip. On the growth chip, pillars ~5μm tall were fabricated using electron-beam lithography and plasma-enhanced dry etching on a double side polished quartz wafer. Mo/Fe catalyst was deposited on top of the pillars and nanotubes were grown by chemical vapor deposition. In approximately 10% of devices, a single tube spanned between the pillars. On the device chip, a silicon substrate covered with 1.9 μm thermal oxide was dry etched to create a mesa ~ 1μm tall. A 5/10 nm Ti/Au gate layer was deposited on the mesa, followed by atomic layer deposition of 60 nm $Al_2O_3$ as gate insulator. On top of that, a 5/80 nm Ti/Au layer was deposited for the contacts. A contact aligner (Karl Suss MJB-3) was used to transfer the nanotube from the growth chip to the device chip with a transfer success rate close to 100%. First measurements (Fig. 1c) were carried out in a 4 K Helium dewar, EDSR was measured at 260 mK, and other measurements were performed in a dilution refrigerator with a base temperature of 100 mK. For the measurement of Fig. 4f, the applied microwave power was -41 dBm and $V_{SD}$ = 5 mV.

## Acknowledgement

We thank Z. Zhong for the initial discussions about the stamping technique, as well as S. M. Frolov, A. Beukman, and J. W. G. van den Berg for valuable suggestions. This research was supported by the Dutch Organization for Fundamental Research on Matter (FOM) and the Netherlands Organization for Scientific Research (NWO).

## Author contributions

F.P. fabricated the devices. F.P. and E.A.L. performed the experiments. L.P.K. supervised the project. F.P., E.A.L. and L.P.K. prepared the manuscript. All authors discussed the results and commented on the manuscript.

## Additional information

Supplementary information accompanies this paper on www.nature.com/naturenanotechnology.

Reprints and permissions information is available online at

http://npg.nature.com/reprintsandpermissions. Correspondence and requests for materials should be addressed to L.P.K. The authors declare that they have no competing financial interests.

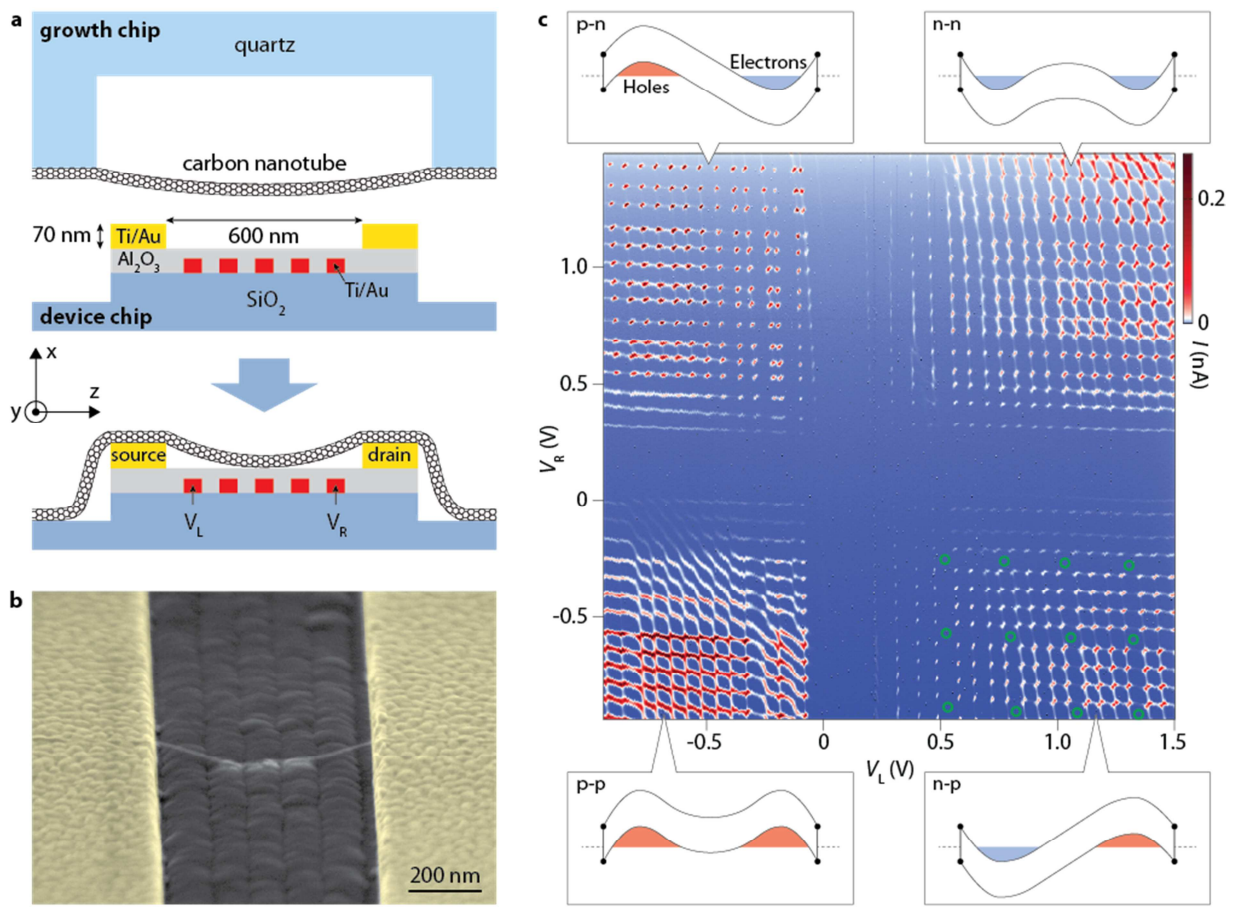

**Figure 1 | A carbon nanotube double quantum dot fabricated by stamping**

**a**, Device schematic and stamping technique. A suspended carbon nanotube is grown between pillars on a transparent quartz substrate. This growth chip is optically aligned over the device chip and brought into contact, so that the nanotube bridges source and drain contacts (yellow). Five local gates (red) embedded in an insulating layer control the electron/hole number and the tunnel rates of the double quantum dot. **b**, Scanning electron micrograph (taken at 75 º inclination) of a device similar to the measured device, showing a single nanotube with a bend. **c**, Current as a function of the left and right gate voltages $V_L$ and $V_R$ at source-drain bias $V_{SD}$ = 1 mV. Regularly spaced Coulomb peaks show clear double quantum dot transport behavior over a large gate space. Individual electrons (holes) are added in n or p-type quantum dots depending on gate settings. Schematic energy diagrams illustrate four types of double quantum dot configuration. Characteristic fourfold periodicity of addition energy is visible revealing the filling of four-electron shells. Green circles in the n-p region indicate double quantum dot with filled shells in both dots.

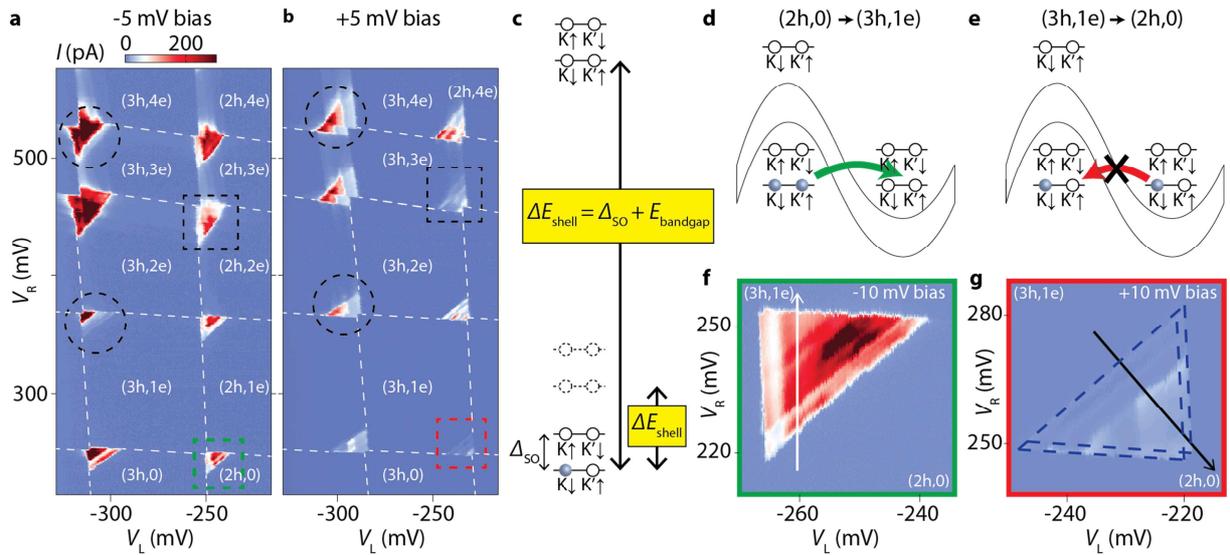

**Figure 2 | Valley-spin blockade in a p-n double quantum dot.**

**a**,**b**, Current at triple point bias triangles in a few-charge p-n double quantum dot at negative (**a**) and positive (**b**) bias. The triangles marked by dashed circles (squares) show a suppressed current at the baseline in **a** (**b**) compared with the same triangles in **b** (**a**). Numbers in brackets denote the occupation of left and right dot with h (e) indicating holes (electrons). **c**, Energy levels of the first shells of electrons and holes. Spin-orbit coupling splits a shell of four electrons (blue) or holes (white) into two levels with energy difference $Δ_{SO}$. A particularly large shell spacing $ΔE_{shell}$ separates the first hole shell from the first electron shell across the bandgap. Dashed levels indicate the nearest higher shell if there would be no bandgap. **d**,**e**, Schematics of the energy levels for the transitions between (2h,0) and (3h,1e). For the non-blockaded bias direction (**d**), either of the two electrons in the left dot can tunnel to the empty shell in the right dot. For the blockaded bias direction (**e**), electron with state $K↓$ in the right dot cannot tunnel to the left dot due to valley-spin conservation. **f**, Measured current for the non-blockade bias direction illustrated in **d**. **g**, Current measured for the opposite bias direction. (Dashed lines mark the position of the triangles). Suppressed current (compared with **f**) across the entire triangles is the signature of valley-spin blockade. **f** and **g** correspond to the square regions in **a** and **b** with same colors.

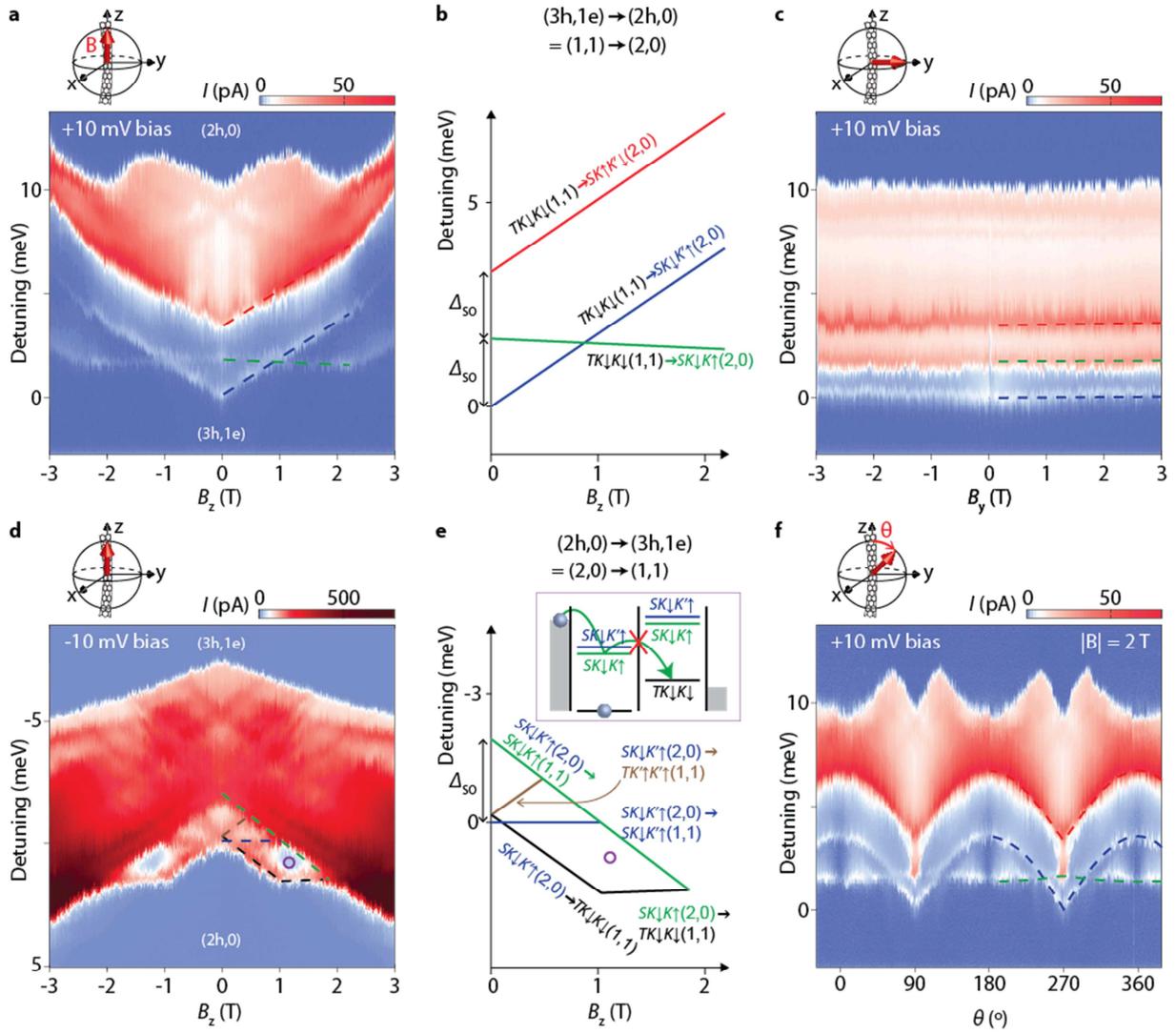

**Figure 3 | Magnetospectroscopy of double quantum dot with valley-spin blockade.**

**a**, Leakage current through valley-spin blockade as a function of detuning (defined by the black arrow in Fig. 2g) and magnetic field $B_z$ along the z-axis. Note that color scale is strongly enhanced compared to Fig. 2g. **b**, The calculated transitions (also marked by dashed lines in **a**), assuming $\Delta_{SO}$ = 1.6 meV and $\mu_{orb}$ = 0.9 meV/T. The two-electron state is denoted as

$S(T)v_1 s_1 v_2 s_2 = (|v_1 s_1 v_2 s_2\rangle - (+)|v_2 s_2 v_1 s_1\rangle)n$ ($n$, normalization factor). **c**, as **a** but with magnetic field $B_y$ along the y-axis. The slope of the transitions (dashes lines) differs from **a**, arising from the g-factor anisotropy. **d**, same as **a**, but for the opposite bias direction. Detuning is defined by the white arrow in Fig. 2f. **e**, The calculated transitions, marked by dashed lines in **d**. Inset shows schematic of valley-spin blockade at the position marked by the purple circle. **f**, Measurement at constant magnetic field $|\mathbf{B}|$ = 2 T while varying the angle $\theta$ between the z-axis and **B** in the z-y plane. Dashed curves are the calculated transitions with colors match to dashed lines in **a** and **c**.

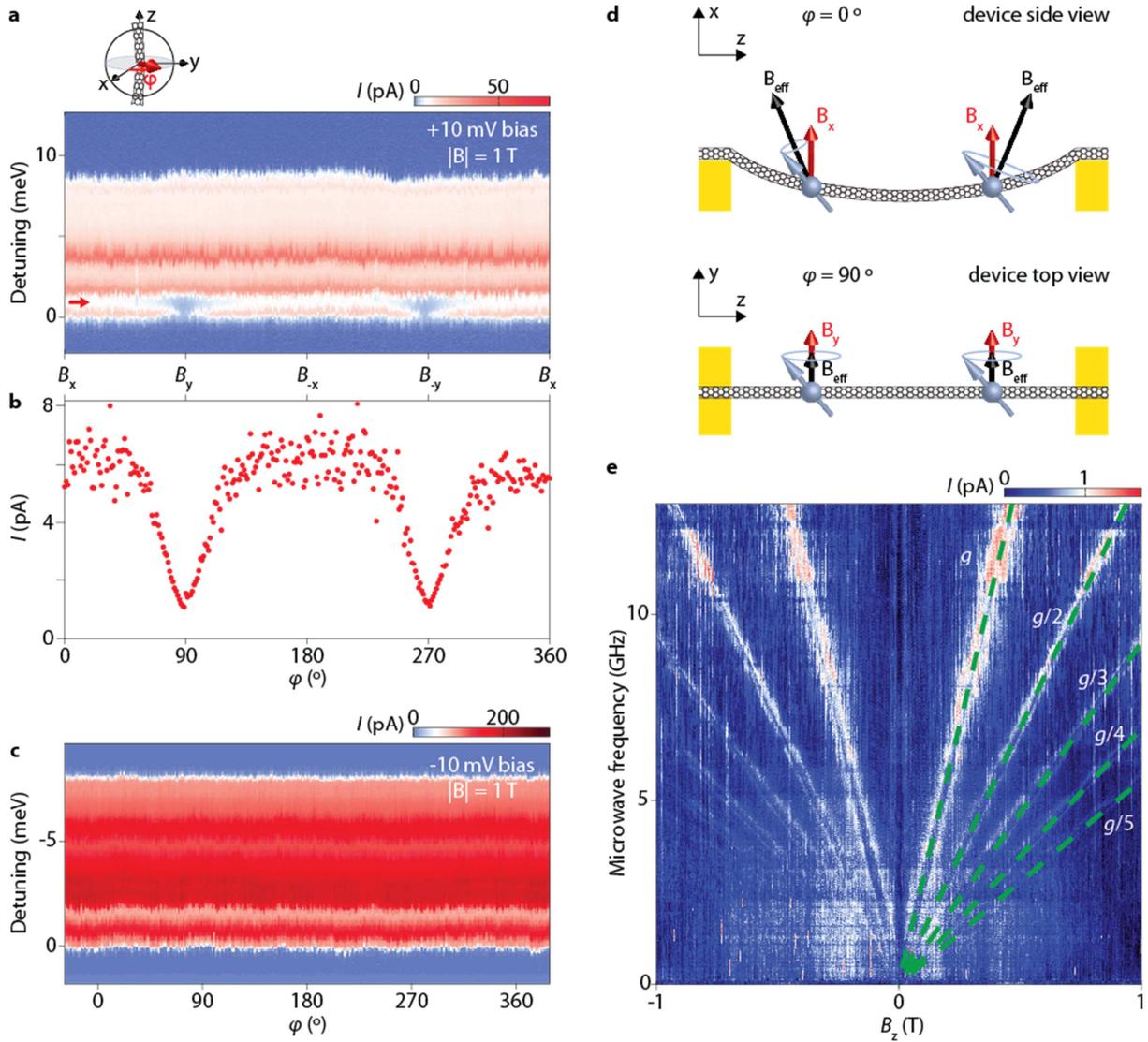

**Figure 4 | The effect of the bend.**

**a**, Current at constant magnetic field |**B**| = 1 T applied in the x-y plane as a function of detuning and of field angle $\varphi$. (Detuning axis as in Fig. 2g.) **b**, Cut along $\varphi$ at the detuning marked by the arrow in **a**. **c**, In the non-blockaded bias direction the current is isotropic. **d**, Schematics for $\varphi = 0°$ and $\varphi = 90°$. When **B**$_{eff}$ varies across the bend, precession of effective spin states about different axes leads to partial lifting of valley-spin blockade. **e**, Current as a function of microwave frequency and $B_z$. V-shaped lines with $g = 2$ are electric dipole spin resonances (EDSR). (To make the resonance clearer, the mean current at each frequency is subtracted.) The resonance lines with slope $g / n$ arise from n-photon transitions[27,29]. EDSR is measured at a blockade transition in the many-electron regime of a second device with similar geometry.

# Supplementary Information

# Valley-spin blockade and spin resonance in carbon nanotubes

Fei Pei, Edward A. Laird, Gary A. Steele, Leo P. Kouwenhoven*

*To whom correspondence should be addressed. E-mail: l.p.kouwenhoven@tudelft.nl

*Kavli Institute of NanoScience, Delft University of Technology,
PO Box 5046, 2600 GA, Delft, The Netherlands*

# Contents

## 1. Energy levels of a two-electron double quantum dot

To understand the transitions observed in Fig. 3 and the anisotropy in Fig. 4, we apply a model incorporating valley and spin energy levels, electrostatic energy, and Coulomb interaction effects.

### 1.1 Single-particle spectrum in a single quantum dot

The low-energy single-particle spectrum of an infinitely long nanotube in an applied magnetic field **B** results from the sum of spin energies, orbital energies and spin-orbit coupling. Neglecting disorder, the Hamiltonian is[1]:

$$H = \tfrac{1}{2} g_s \, \mu_B \, \mathbf{s}\cdot\mathbf{B} + v_3 \, \mu_{\text{orb}} \, \mathbf{B}\cdot\hat{\mathbf{z}} + \left(\tfrac{1}{2} \Delta_{\text{SO}} \, v_3 \, \mathbf{s}\cdot\hat{\mathbf{z}}\right), \qquad (1)$$

$v_i$ and $s_i$ are the Pauli matrices in respectively valley ($K$, $K'$) and spin ($\downarrow$, $\uparrow$) degrees of freedom, $\mu_B$ is the Bohr magneton, $g_s = 2$ is the electron spin $g$-factor, and $\hat{\mathbf{z}}$ is the unit vector along the nanotube axis. With the magnetic field $B_z$ applied along a straight nanotube, the single-particle energies are:

$$E_{\text{single-particle}}(v, s, B_z) = \tfrac{1}{2} g_s \, \mu_B \, s \, B_z + v \, \mu_{\text{orb}} \, B_z + \left(\tfrac{1}{2} \Delta_{\text{SO}} \, v \, s\right), \qquad (2)$$

$v = +1$ (-1) corresponds to valley state $K'$ ($K$) and $s = +1$ (-1) corresponds to spin state $\downarrow$ ($\uparrow$).

A nanotube of finite length becomes a quantum dot with quantization of the longitudinal momentum. Each quantized longitudinal mode is called a *shell* and contains four states, because valley and spin each take two values. Fig. S1a shows the $B_z$-dependence of the single-particle energies within a single shell (ignoring charging energies). The energy difference between two consecutive shells is defined as the shell level spacing $\Delta E_{shell}$.

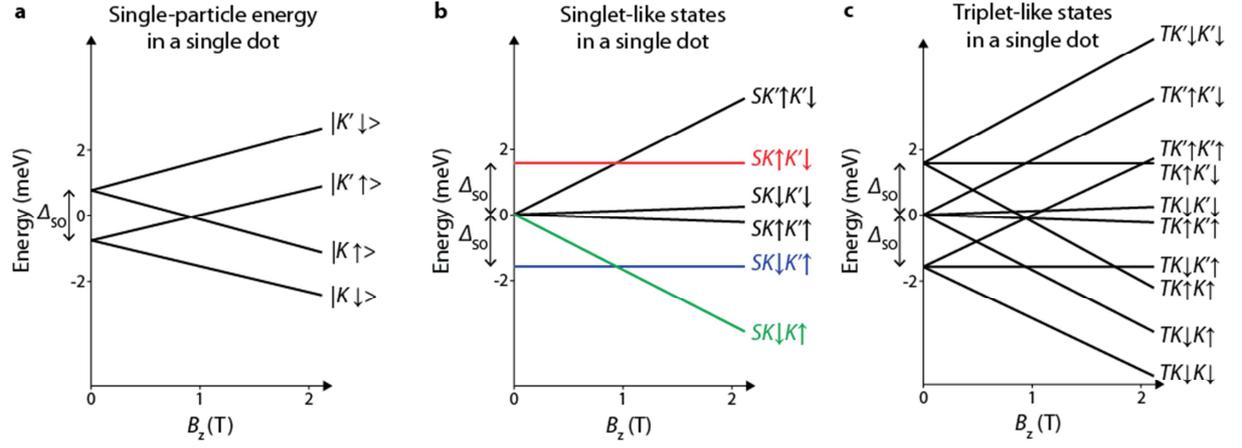

**Figure S1 | Single-particle and two-electron energies in a single quantum dot. a**, Single-particle energy spectrum of a single shell versus $B_z$ parallel to the nanotube. At $B_z = 0$, spin-orbit coupling creates two doublets with energy difference $\Delta_{SO}$. **b,c**, Single dot two-electron energy levels as a function of $B_z$. Valley-spin antisymmetric (singlet-like) states are plotted in **b**. (The energy scale is offset from that in **a** by the charging energy $E_C$.) Valley-spin symmetric (triplet-like) states are plotted in **c**. (The levels are offset by the sum of $E_C$ and $E'_{ST}$.) The levels are calculated using the experimental values $\Delta_{SO} = 1.6$ meV and $\mu_{orb} = 0.9$ meV/T.

### 1.2 Two-electron spectrum in a single quantum dot

The two-electron wave function is the product of a valley-spin part and an orbital part:

$$\Psi_{\text{two-electron}} = \Psi_{\text{valley-spin}}(v_1, s_1, v_2, s_2) \otimes \Psi_{\text{orbital}}, \qquad (3)$$

where $v_1$, $s_1$, $v_2$, and $s_2$ are the valley and spin quantum numbers of the two electrons. The Pauli exclusion principle requires that the total wave function is antisymmetric with respect to electron exchange. Therefore an antisymmetric (symmetric) valley-spin part implies a symmetric (antisymmetric) orbital part.

In the absence of valley freedom, $\Psi_{\text{spin}}(s_1, s_2)$ replaces $\Psi_{\text{valley-spin}}(v_1, s_1, v_2, s_2)$ in equation (3). An antisymmetric spin state, i.e. a *singlet* state, can be constructed with both electrons occupying the same orbital state. In contrast, for the *triplet* states with a symmetric spin part the two electrons need to occupy different orbital states in order to satisfy Pauli's exclusion principle. This increases the energy of the triplet states (degenerate at $B = 0$) above the energy of the singlet state by an amount denoted as $E_{ST}$ (being equal to the orbital spacing minus the difference in Coulomb repulsion between two electrons in a singlet versus a triplet state).

In our case incorporating valley freedom, we refer to the two-electron states with antisymmetric (symmetric) valley-spin parts as singlet-like (triplet-like) states and denote them by *S* (*T*). We can now construct six singlet-like states, occupying valleys within the same shell. In order to construct the triplet-like states we need to include at least two shells, resulting in total ten triplet-like states[2-4]. These are summarized in Table S1.

**Table S1 | Two-electron states comparison between spin-only systems and carbon nanotubes.**

| Spin-only systems | | | |
|---|---|---|---|
| $\Psi_{\text{spin}}$ | $\Psi_{\text{orbital}}$ | Referred to as | Number of states |
| Antisymmetric | Symmetric | Singlet | 1 |
| Symmetric | Antisymmetric | Triplet | 3 |
| **Carbon nanotubes** | | | |
| $\Psi_{\text{valley-spin}}$ | $\Psi_{\text{orbital}}$ | Referred to as | Number of states |
| Antisymmetric | Symmetric | Singlet-like | 6 |
| Symmetric | Antisymmetric | Triplet-like | 10 |

The energies of the two-electron states can be calculated using the constant interaction model[5]. The energies of the singlet-like states are taken as the sum of the single-particle energies (equation. (2)) of the two electrons $E(v_1, s_1)$ and $E(v_2, s_2)$, plus a charging energy $E_C$:

$$E_{\text{singlet-like}} = E(v_1, s_1) + E(v_2, s_2) + E_C . \qquad (4)$$

Singlet-like states consist of two electrons from the same shell, while electrons from two shells are required to form a triplet-like state. This increases the energy of the triplet-like states by $E'_{ST}$:

$$E_{\text{triplet-like}} = E(v_1, s_1) + E(v_2, s_2) + E_C + E'_{ST} . \qquad (5)$$

The allowed combination of the valley-spin quantum numbers for the singlet-like and triplet-like states are listed in Table S2. The other combinations of the valley-spin quantum numbers lead to states that either differ only by a phase factor, or are unphysical states equal to zero.

A sufficiently large $E'_{ST}$ is crucial for obtaining Pauli blockade in carbon nanotubes. The complication is that $E'_{ST}$ is not just equal to the shell spacing $\Delta E_{\text{shell}}$, but is instead reduced by the energy term $C$ resulting from Coulomb interaction effects[5]:

$$E'_{ST} = \Delta E_{\text{shell}} - C . \qquad (6)$$

In a suspended nanotube, Coulomb interaction effects become particularly important compared to those in quantum dots made in III-V materials, because there is no surrounding dielectric to screen them. To overcome this problem, we configure the device in such a way that the nearest higher shell in the left dot is across the bandgap. This significantly increases $E'_{ST}$ since $\Delta E_{\text{shell}}$ includes the bandgap. Therefore the triplet-like states are much higher in energy than the singlet-like states. In Fig. S1b (c) we plot the magnetic field dependence of the singlet-like (triplet-like) states.

**Table S2 | Energies of the singlet-like and triplet-like states at zero magnetic field.**

| **Singlet-like states**, $Sv_1s_1v_2s_2 = (|v_1s_1v_2s_2\rangle - |v_2s_2v_1s_1\rangle)/\sqrt{2}$ | | |
|---|---|---|
| $E = -\Delta_{SO} + E_C$ | $E = E_C$ | $E = \Delta_{SO} + E_C$ |
| SK↓K'↑ | SK↓K↑ | SK↑K'↓ |
| | SK↑K'↑ | |
| | SK↓K'↓ | |
| | SK'↑K'↓ | |
| **Triplet-like states**, $Tv_1s_1v_2s_2 = (|v_1s_1v_2s_2\rangle + |v_2s_2v_1s_1\rangle)n$, where *n* is the normalization factor | | |
| $E = -\Delta_{SO} + E_C + E'_{ST}$ | $E = E_C + E'_{ST}$ | $E = \Delta_{SO} + E_C + E'_{ST}$ |
| TK↓K↓ | TK↓K↑ | TK↑K↑ |
| TK↓K'↑ | TK↑K'↑ | TK↑K'↓ |
| TK'↑K'↑ | TK↓K'↓ | TK'↓K'↓ |
| | TK'↑K'↓ | |

## 1.3 Two-electron spectrum in a double quantum dot

Two charge configurations are relevant for the transitions in the main text: (1,1) and (2,0). For a weakly coupled double dot, the singlet-like and triplet-like energies can be calculated as the sum of the single dot energies and the electrostatic term $E_{es}$:

$$E_{singlet-like} = E(v_1, s_1) + E(v_2, s_2) + E_C + E_{es}, \qquad (7)$$

$$E_{triplet-like} = E(v_1, s_1) + E(v_2, s_2) + E_C + E'_{ST} + E_{es}, \qquad (8)$$

where

$$E'_{ST} = \begin{cases} \varepsilon_{11}, & \text{for (1,1) charge states} \\ \varepsilon_{20}, & \text{for (2,0) charge states.} \end{cases} \qquad (9)$$

The electrostatic energy is set by the gate voltages and raises the energy of the (1,1) configuration by an amount called detuning, $\Delta$:

$$E_{es} = \begin{cases} 0, & \text{for (2,0) charge states} \\ \Delta, & \text{for (1,1) charge states.} \end{cases} \qquad (10)$$

For the (1,1) → (2,0) transition, we study the charge cycle (1,0) → (1,1) → (2,0) → (1,0), and assume that the (1,1) state is in the ground state, *TK↓K↓*. By varying the detuning, *TK↓K↓*(1,1) is shifted along the *S*(2,0) spectrum. Every time it aligns with a particular *S*(2,0) state, electrons can tunnel to the left dot and contribute to the current. Fig. S2 illustrates this spectroscopy of *TK↓K↓*(1,1) → *S*(2,0) and shows the three transitions that we have observed in the measurements.

For opposite bias, we study the reverse charge cycle including the (2,0) → (1,1) transition. We use the (2,0) ground states to probe the spectrum of the sixteen (1,1) states (Fig. S3a). For $B_z < 1$ T, the initial ground state is *SK↓K'↑*(2,0). The calculated transition detunings from this state as a function of $B_z$ are shown in Fig. S3b. At higher field, the system undergoes a ground state crossing, with *SK↓K↑*(2,0) becoming the ground state for $B_z > 1$ T. The transition detunings are shown in Fig. S3c.

To identify states participating in transport, we compare the data in Fig. 3d to the calculated transitions in Fig. S3b and c. The identified transitions are shown in Fig. S3d. (Transitions at higher detuning could not be clearly identified because of their broader linewidths.) Comparison with the data yields $\varepsilon_{11} = 180\ \mu\mathrm{eV}$.

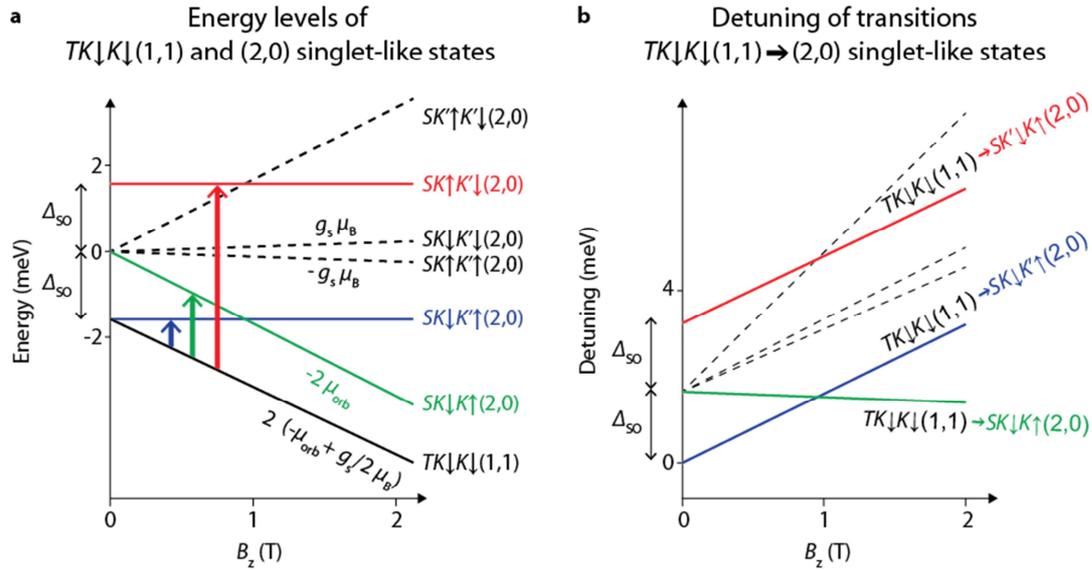

**Figure S2 | Energy levels for the (1,1) → (2,0) transition. a**, Energy levels relevant for the (1,1) → (2,0) transitions. The lowest energy state in the (1,1) configuration is $TK{\downarrow}K{\downarrow}$ for a finite $B_z$. This is used to probe the (2,0) spectrum. (The levels are offset by the charging energy $E_C$.) The zero of detuning is defined by the alignment of $TK{\downarrow}K{\downarrow}(1,1)$ with $SK{\downarrow}K'{\uparrow}(2,0)$ at zero field. The slope of the lines is marked for selected states. A positive detuning shifts up the energy of $TK{\downarrow}K{\downarrow}(1,1)$. A transition is possible when the energies of the (1,1) and (2,0) states align. **b**, Detunings required for the possible transitions as a function of $B_z$. Colored transition lines are the observed ones in the measurements (also marked by the arrows in the same color in **a**). Dashed transitions are not observed. The conversion from gate voltage to energy is obtained from the size of the bias triangle.

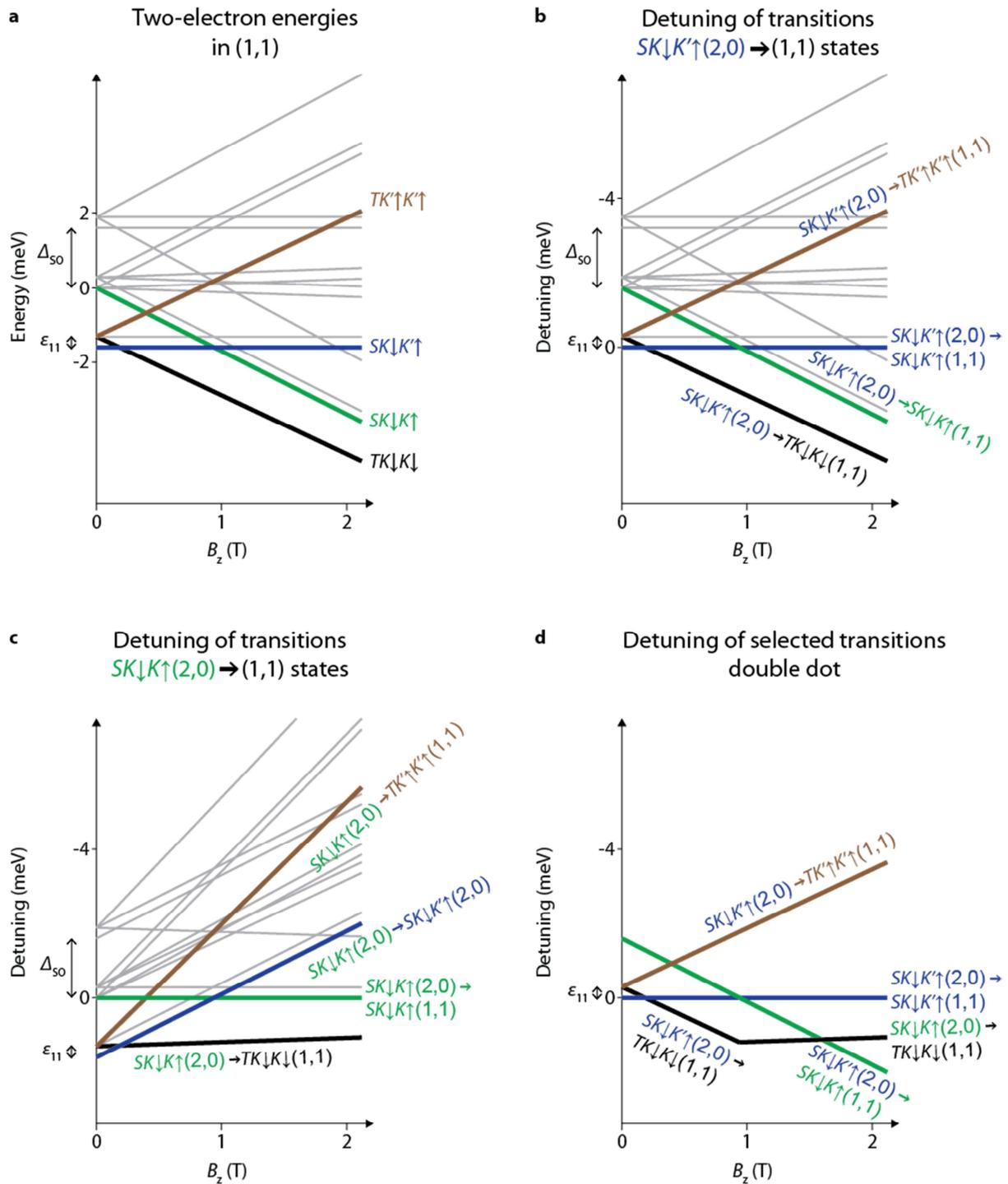

**Figure S3 | Energy levels for the (2,0) → (1,1) transition. a**, Two-electron energy levels of the (1,1) charge states as a function of $B_z$. (The levels are offset by the charging energy $E_C$.) Of the sixteen states, four (in color) are important for our measurements. The energy difference between $TK{\downarrow}K{\downarrow}$ and $SK{\downarrow}K'{\uparrow}$ at zero field is defined as $\varepsilon_{11}$. **b**, Detunings as a function of $B_z$ for the possible transitions from the initial state $SK{\downarrow}K'{\uparrow}(2,0)$ to the sixteen (1,1) states. The zero of detuning is defined by the alignment of $SK{\downarrow}K'{\uparrow}(2,0)$ with $SK{\downarrow}K'{\uparrow}(1,1)$. **c**, Same as **b**, but with the initial state $SK{\downarrow}K{\uparrow}(2,0)$. **d**, Detuning as a function of $B_z$ for the identified transitions in Fig. 3d. $SK{\downarrow}K{\uparrow}$ takes over from $SK{\downarrow}K'{\uparrow}$ as the (2,0) ground state at field ~ 1 T, resulting in a kink for the lowest transitions (black).

## 1.4 Including disorder

To obtain an optimal comparison with the transitions in Fig. 3c and f, we incorporate the valley-mixing term $\Delta_{KK'}$ into equation (1)[1]. Defining $\theta$ as the angle between the magnetic field and the nanotube axis and using $K'\downarrow$, $K\uparrow$, $K'\uparrow$, $K\downarrow$ as the basis, the Hamiltonian becomes[6]:

$$H = \frac{1}{2} g_s \mu_B B \begin{pmatrix} \cos\theta & 0 & \sin\theta & 0 \\ 0 & -\cos\theta & 0 & \sin\theta \\ \sin\theta & 0 & -\cos\theta & 0 \\ 0 & \sin\theta & 0 & \cos\theta \end{pmatrix} + \mu_{orb} B \cos\theta \begin{pmatrix} 1 & 0 & 0 & 0 \\ 0 & -1 & 0 & 0 \\ 0 & 0 & 1 & 0 \\ 0 & 0 & 0 & -1 \end{pmatrix}$$
$$+ \frac{1}{2} \begin{pmatrix} \Delta_{SO} & 0 & 0 & \Delta_{KK'} \\ 0 & \Delta_{SO} & \Delta_{KK'} & 0 \\ 0 & \Delta_{KK'} & -\Delta_{SO} & 0 \\ \Delta_{KK'} & 0 & 0 & -\Delta_{SO} \end{pmatrix} \qquad (11)$$

The single-particle energies are obtained by numerically diagonalizing the Hamiltonian of equation (11). The calculated curves in Fig. 3c and f are then obtained following the same procedure as described in the previous section. $\Delta_{KK'}$ is extracted from the data described in Section 6.

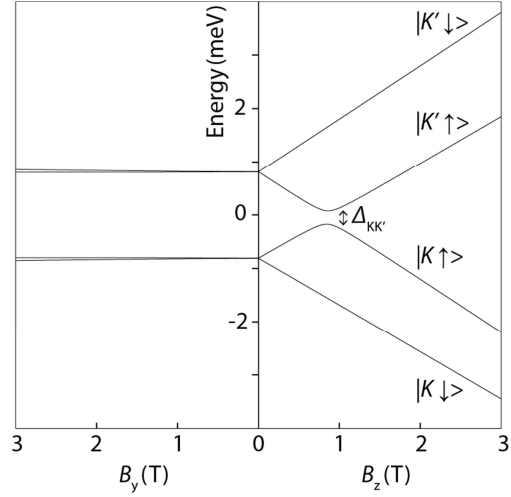

**Figure S4 | Single-particle energies including disorder.** The levels are calculated using equation (11) with the experimental values $\Delta_{SO} = 1.6$ meV, $\Delta_{KK'} = 0.25$ meV, and $\mu_{orb} = 0.9$ meV/T.

## 2. Selection rules for valley-spin relaxation

To investigate the selection rules for the observed valley-spin relaxation transitions, we discuss here several mechanisms that can cause valley-spin relaxation: spin-orbit interaction[1,7,8,9], intervalley scattering[10], and hyperfine interaction with the nuclei[8].

The six possible (1,1) → (2,0) transitions are listed in Table S3. Note that the valley-spin symmetry of the two-electron state is not conserved for these transitions. For the opposite bias direction, we only list the observed and identified transitions for clarity (Table S4). Three out of five transitions involve a change of symmetry. Several other transitions are visible for higher detunings in Fig, 3d, but could not be identified because they overlap. Therefore (2,0) → (1,1) transitions that are not listed in Table S3 cannot be classified as not observed.

**Table S3 | Selective valley-spin relaxation for the (1,1) → (2,0) transitions.**

| (1,1) state | (2,0) state | Valley-spin symmetry change | Number of valley-flips | Number of spin-flips | Transition observed |
|---|---|---|---|---|---|
| **TK↓K↓** | **SK↓K'↑** | yes | **1** | **1** | **yes** |
| **TK↓K↓** | **SK↓K↑** | yes | **0** | **1** | **yes** |
| TK↓K↓ | SK↑K'↑ | yes | 1 | 2 | no |
| TK↓K↓ | SK↓K'↓ | yes | 1 | 0 | no |
| TK↓K↓ | SK'↑K'↓ | yes | 2 | 1 | no |
| **TK↓K↓** | **SK↑K'↓** | yes | **1** | **1** | **yes** |

**Table S4 | Observed and identified (2,0) → (1,1) transitions.**

| (2,0) state | (1,1) state | Valley-spin symmetry change | Number of valley-flips | Number of spin-flips |
|---|---|---|---|---|
| **SK↓K'↑** | **SK↓K'↑** | no | **0** | **0** |
| **SK↓K'↑** | **SK↓K↑** | no | **1** | **0** |
| **SK↓K'↑** | **TK↓K↓** | yes | **1** | **1** |
| **SK↓K'↑** | **TK'↑K'↑** | yes | **1** | **1** |
| **SK↓K↑** | **TK↓K↓** | yes | **0** | **1** |

Valley-spin relaxation can be attributed to spin-orbit interaction[8,11-13] through the mechanisms of bend-mediated relaxation[1] as explained in the main text or spin-phonon coupling[7,9]. Alternatively, hyperfine coupling can also mix both valley and spin states[14]. Valley mixing can arise from the disorder[10] (presumably responsible for the observed transition $SK\downarrow K'\uparrow(2,0) \rightarrow SK\downarrow K\uparrow(1,1)$). However valley mixing alone cannot be responsible for the other observed transitions which include a spin-flip. We therefore suggest that spin-orbit interaction and hyperfine coupling contribute to the most of the observed and identified transitions. However, the corresponding relaxation rates and observed leakage currents are not understood.

## 3. Valley-spin relaxation around zero magnetic field

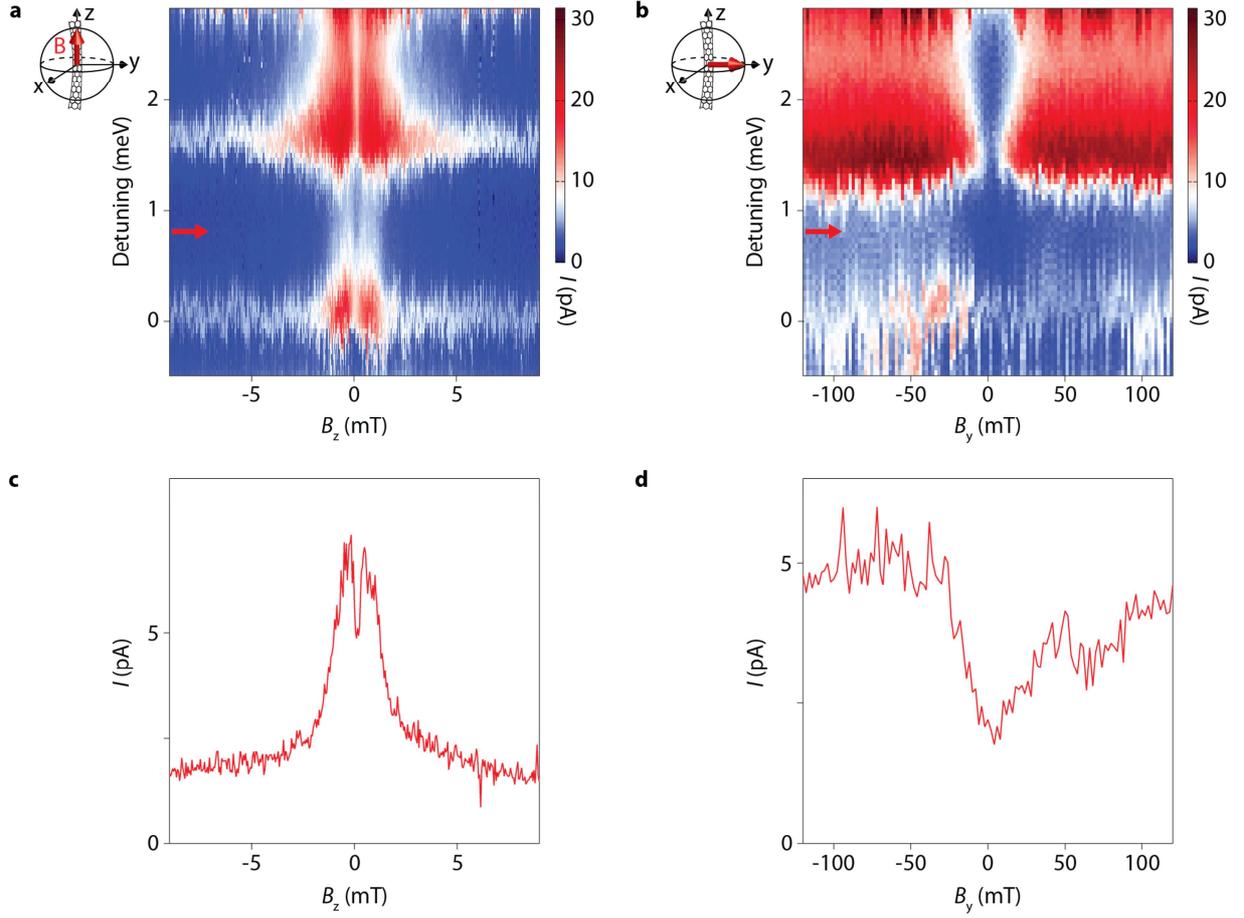

**Figure S5 | Contrasting valley-spin relaxation for parallel and perpendicular magnetic field. a,b,** Valley-spin leakage current around $B_z = 0$ (**a**) and $B_y = 0$ (**b**). Detuning axis is the same as in Fig. 3a. **c,d,** Cuts along $B_z$ (**c**) and $B_y$ (**d**) at the detunings marked by the arrows in **a** and **b**.

The behavior of the leakage current at low magnetic field can give insight into the electron spin relaxation mechanisms[8,15]. A striking difference in the leakage current is observed for different field directions (Fig. S5). With the field applied along the nanotube, we see a split peak in the current around zero field (Fig. S5a), whereas for perpendicular field, a broad dip is observed (Fig. S5b).

A current peak at zero field was previously measured in an isotopically enriched 99% $^{13}$C nanotube device, and attributed to an enhanced hyperfine interaction, although with a surprisingly large hyperfine coupling constant deduced from the peak width[8]. In our case, the peak full width at half maximum is approximately ten times smaller. Since this peak width scales with the square root of the $^{13}$C concentration, this corroborates the hyperfine coupling constant measured in ref. 8, assuming the same quantum dot size and nanotube diameter. The splitting of the peak could be due to exchange[15], although in this case the size of the splitting would be expected to depend on detuning, in contrast to what is observed. A local current minimum at $B_z = 0$ can arise due to spin-orbit coupling as described in ref. 10. A split peak due to hyperfine interaction is also predicted in ref. 14, although that theory does not include spin-orbit coupling.

The current minimum at zero perpendicular field is likely due to the influence of spin-orbit coupling, which can cause spin relaxation only at finite field[8,11-13,16]. Alternatively, such a minimum has been predicted to arise from disorder-induced valley mixing[10]. The different behavior in parallel and perpendicular fields presumably reflects anisotropy of either the spin-orbit or hyperfine coupling[17]. The difference in the zero field current between Fig. S5c and d can arise from a small magnetic field offset in the setup.

## 4. The effect of the bend

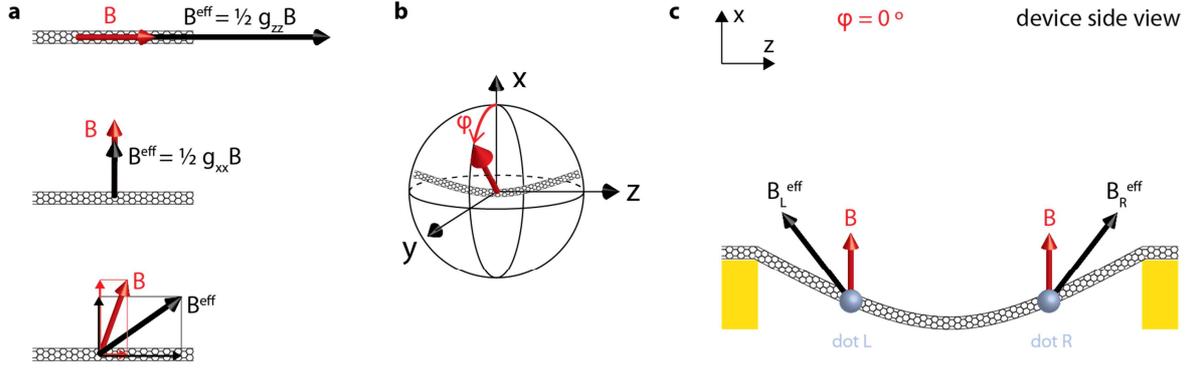

**Figure S6 | Schematic of magnetic field applied to a bent nanotube. a**, Magnetic field (red arrow) applied at different angles, and the resulting effective magnetic field (black arrow) acts on the combined valley-spin states. **b**, Coordinate system for our bent nanotube. Magnetic field is applied in the x-y plane. **c**, $\mathbf{B}_{eff}$ varies across the bend in the case of a bent nanotube with the field applied along the x-axis.

In this section we elaborate on the effective magnetic field in a bent nanotube mentioned in the main text. The two single-particle states ($K'\uparrow$, $K\downarrow$) that comprise the low-energy doublet form an effective spin-1/2 subspace. Within this subspace, an effective Hamiltonian derived from equation (1) is written[1]:

$$H^* = \frac{1}{2}\mu_B\, \mathbf{s}^* \cdot \mathbf{g} \cdot \mathbf{B}, \tag{12}$$

where $\mathbf{s}^*$ is a vector of effective spin Pauli matrices. The **g**-tensor is given by:

$$\mathbf{g} = \begin{pmatrix} g_{xx} & 0 & 0 \\ 0 & g_{yy} & 0 \\ 0 & 0 & g_{zz} \end{pmatrix}, \tag{13}$$

where,

$$g_{xx} = g_{yy} = \frac{g_s \Delta_{KK'}}{\sqrt{\Delta_{KK'}^2 + \Delta_{SO}^2}}, \tag{14}$$

$$g_{zz} = g_s + \frac{2\mu_{orb}\Delta_{SO}}{\mu_B \sqrt{\Delta_{KK'}^2 + \Delta_{SO}^2}}. \tag{15}$$

For a straight nanotube pointing along the z-axis, $g_{xx} = g_{yy}$ due to the rotational symmetry of the nanotube. In the case of a ultra-clean nanotube, $g_{yy}$ approaches zero. In a disorder-dominated nanotube, where $\Delta_{KK'} \gg \Delta_{SO}$, $g_{yy} \sim 2$. For our nanotube with $\Delta_{KK'}$ determined in Section 6, $g_{yy} = 0.2$.

Within the ($K'\uparrow$, $K\downarrow$) subspace, the evolution of the electron state is the same as that of a spin-1/2 particle in an effective magnetic field[1] $\mathbf{B}_{eff} = \mathbf{g} \cdot \mathbf{B}/g_s$. Fig. S6a illustrates the magnetic field **B** applied at different angles to the nanotube and the resulting effective field **B**$_{\text{eff}}$. For a parallel **B** (Fig. S6a upper panel), **B**$_{\text{eff}}$ points to the same direction and the magnitude depends on $g_{zz}$. If 1 T is applied parallel to the nanotube, and we take $g_{zz} = 30$, $B_{\text{eff}}$ will be 15 T. This large effective field does not act directly on the electron spins, but on the combined valley-spin states. In the case of a perpendicular **B** (Fig. S6a middle panel), **B**$_{\text{eff}}$ is in the same direction and the magnitude depends on $g_{xx}$. For **B** at other angles (Fig. S6a lower panel), **B**$_{\text{eff}}$ does not point along **B** and can be constructed using vector decomposition. In the case of a bent nanotube with the field applied along the x-axis (Fig. S6c), the angle between the applied field and the local nanotube axis varies across the bend, therefore **B**$_{\text{eff}}$ also varies across the bend.

## 5. Alignment of the magnetic field relative to the nanotube

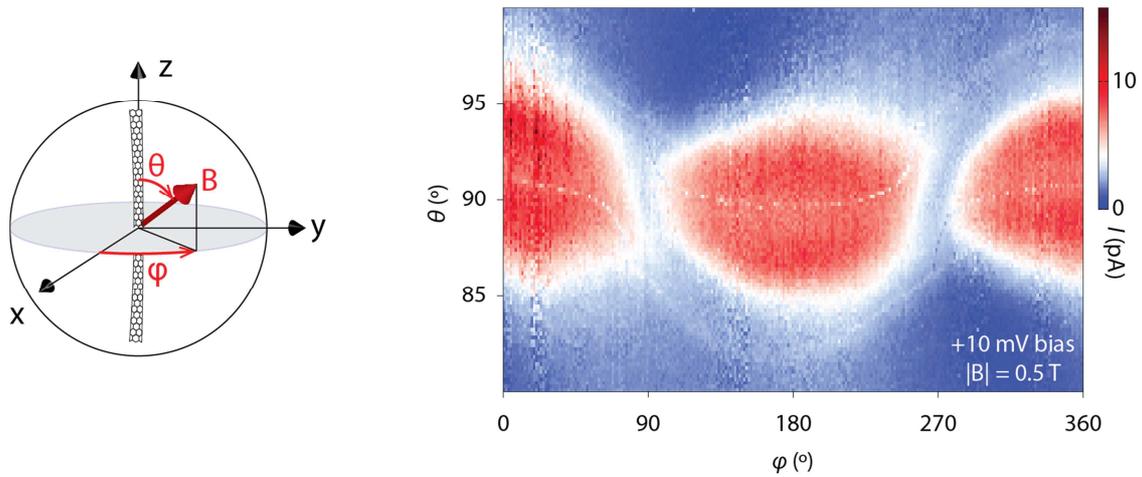

**Figure S7 | Accurate alignment of the magnetic field.** Current near zero detuning (position marked by the red arrow in Fig. 4a) in a constant magnetic field |**B**| = 0.5 T while varying the angle $\varphi$ and $\theta$ close to the plane perpendicular to the nanotube axis. The near-symmetry in both $\varphi$ and $\theta$ confirms our estimation of the nanotube local coordinate system and indicates a good alignment of the applied magnetic field.

Correct alignment of the applied magnetic field is crucial for our analysis of the *g*-factor anisotropy. To maintain ultracleanness of the nanotube, we did not image the device, making the orientation of the z-axis relative to the chip uncertain. The axis of the nanotube is identified by measuring the current anisotropy as in Fig. 3f and Fig. 4a. The nanotube coordinate system is then constructed so that the current is symmetric with respect to field angle about 180°. All our data is taken using the nanotube coordinate system.

The data in Fig. S7 confirms that the anisotropy measured in Fig. 4a is not due to field misalignment; the current minimum near $\varphi$ = 90° is observed even if the field is tilted away from the equator ($\theta$ = 90°).

## 6. Angle independence of valley mixing

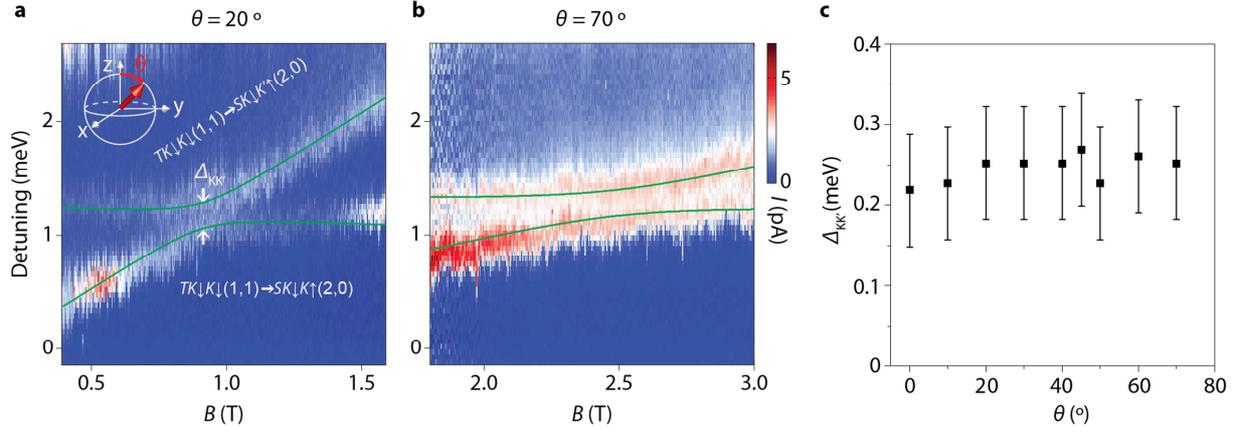

**Figure S8 | Angle independence of disorder-induced valley-mixing. a**, Leakage current as a function of detuning (defined in Fig. 2g) and magnetic field at the angle $\theta = 20°$. Green curves are calculated using equation (16) yielding $\Delta_{KK'} = 0.25 \pm 0.07$ meV. **b**, as **a** but at $\theta = 70°$, and $\Delta_{KK'} = 0.25 \pm 0.07$ meV. **c**, $\Delta_{KK'}$ as a function of $\theta$. The error bars are determined by average linewidth of the transitions.

The disorder-induced valley-mixing strength $\Delta_{KK'}$ can be obtained from magnetospectroscopy[18]. In Fig. S8a, the transitions $TK{\downarrow}K{\downarrow}(1,1) \rightarrow SK{\downarrow}K{\uparrow}(2,0)$ and $TK{\downarrow}K{\downarrow}(1,1) \rightarrow SK{\downarrow}K'{\uparrow}(2,0)$ show an avoided crossing, which arises because the final states differ in the valley of one electron. The splitting of the anticrossing is equal to $\Delta_{KK'}$, and is extracted by comparing the transition detunings to a simple model of two anticrossing energy levels. We approximate the two energy levels involved by the eigenvalues of the following Hamiltonian:

$$H_{\text{crossing}} = \begin{pmatrix} A + (B - B_c)q_1 & \Delta_{KK'}/2 \\ \Delta_{KK'}/2 & A + (B - B_c)q_2 \end{pmatrix}, \tag{16}$$

where the anticrossing field $B_c$, and the slopes $q_1$ and $q_2$ of the two transitions are determined from equation (11). The detuning $A$ at the anticrossing and $\Delta_{KK'}$ are taken as free parameters.

Fig. S8a and b show calculated anticrossings superimposed on the data for two magnetic field directions. Analyses of similar data over a range of field directions (Fig. S8c) show that $\Delta_{KK'} = 0.25 \pm 0.08$ meV is independent of angle.

## 7. EDSR in the first device

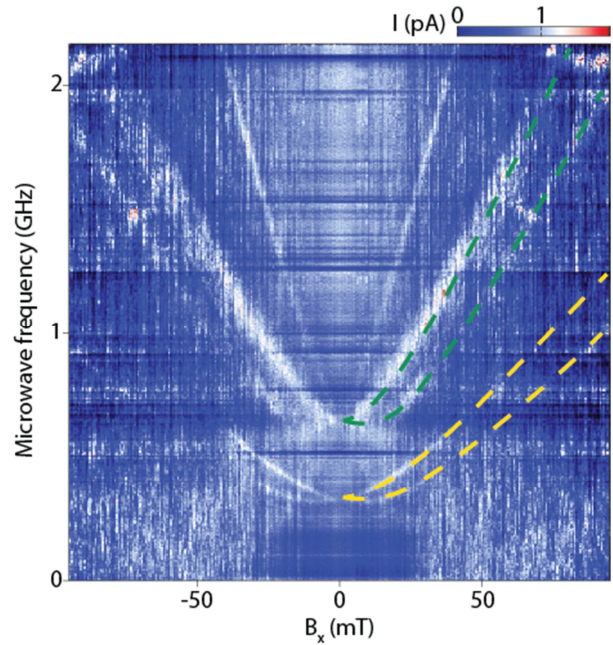

**Figure S9 | EDSR at transition (1h,1e) → (0,0) of the first device.** Current as a function of microwave frequency and $B_x$. Several electric dipole spin resonances are observed. Green curves are guides to the eye indicating the most pronounced resonance lines. Yellow curves indicate subharmonics at half of the frequency. To make the resonances clearer, field and frequency independent background currents have been subtracted. The applied microwave power was -31 dBm and $V_{SD}$ = 5 mV.

Electric dipole spin resonance (EDSR) is also detected in the first device in a different cooldown and at the blockaded transition (1h,1e) → (0,0). As shown in Fig. S9, we observe a complex spectrum with many resonances, which presumably reflects spin-orbit interaction, the bend, and exchange. The green curves are guides to the eye drawn over the two strongest resonance lines, The yellow curves are drawn at half the frequency and overlap two faint resonances suggesting that these arise from two-photon transitions[19-22]. However the full spectrum is not understood.